# Line Clipping in $E^3$ with Expected Complexity O(1)

Václav Skala[1]
Departments of Informatics and Computer Science
University of West Bohemia
Americká 42, Box 314
306 14 Plzeò
Czech Republic
skala@kiv.zcu.cz http://yoyo.zcu.cz/~skala

**Abstract**

A new line clipping algorithm against convex polyhedron in $E^3$ with an expected complexity *O(1)* is presented. The suggested approach is based on two orthogonal projections to $E^2$ co-ordinate system and on pre-processing of the given polyhedron. The pre-processing enables to speed up solution significantly. The proposed method is convenient for those applications when many lines are clipped against constant convex polyhedron. Theoretical considerations and experimental results are also presented.

## 1. Introduction

Pre-processing is mostly used in many applications to speed up problem solutions significantly because it enables us to decrease run time complexity substantially. Many algorithms for line clipping in $E^2$ and in $E^3$ have been developed, see [Ska94a], [Ska96a] for references. Algorithms for line clipping in $E^3$ are mostly based on the Cyrus-Beck (CB) algorithm and its modifications. The aim of the line clipping algorithm is to find a part of the given line *p* which is inside of the given polyhedron *P*, see Fig.1.1. Algorithms for line clipping are mostly restricted to line clipping against convex polyhedron. Because the line clipping problem solution is a bottleneck of many packages and applications it is convenient to use the fastest algorithm even it is of *O(N)* complexity. Algorithm comparisons can be found in [Ska94a], [Ska96a]. Nevertheless there is also a possibility to develop an algorithm with lower complexity, e.g. with complexity $O(\sqrt{N})$ [Ska96c].

[1] Supported by the grant UWB-156/1995

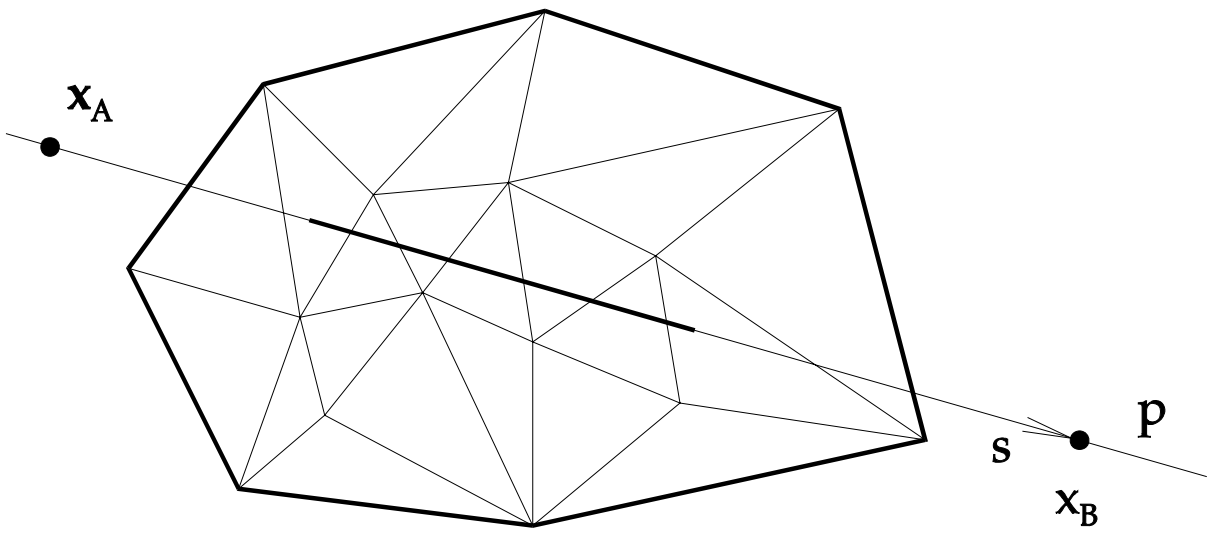

**Line clipping by convex polyhedron in E$^3$**
Figure 1.1

There are applications where the clipping polyhedron is constant for many clipped lines. The known algorithms do not use pre-processing in order to decrease processing time significantly. The proposed method described in this paper is based on the algorithm for two dimensional space clipping with expected *O(1)* complexity, see [Ska96b].

Let us assume that a convex polyhedron *P* is defined by triangular facets (generally it is not necessary). We need to find an **effective test** whether the line *p* intersects a facet of the given polyhedron *P*, see Fig.1.1, and then compute intersection points. Unfortunately the CB algorithm **computes all** intersection points and selects the appropriate two of those. A simple modification of the CB algorithm can speed up the processing, see [Ska96a].

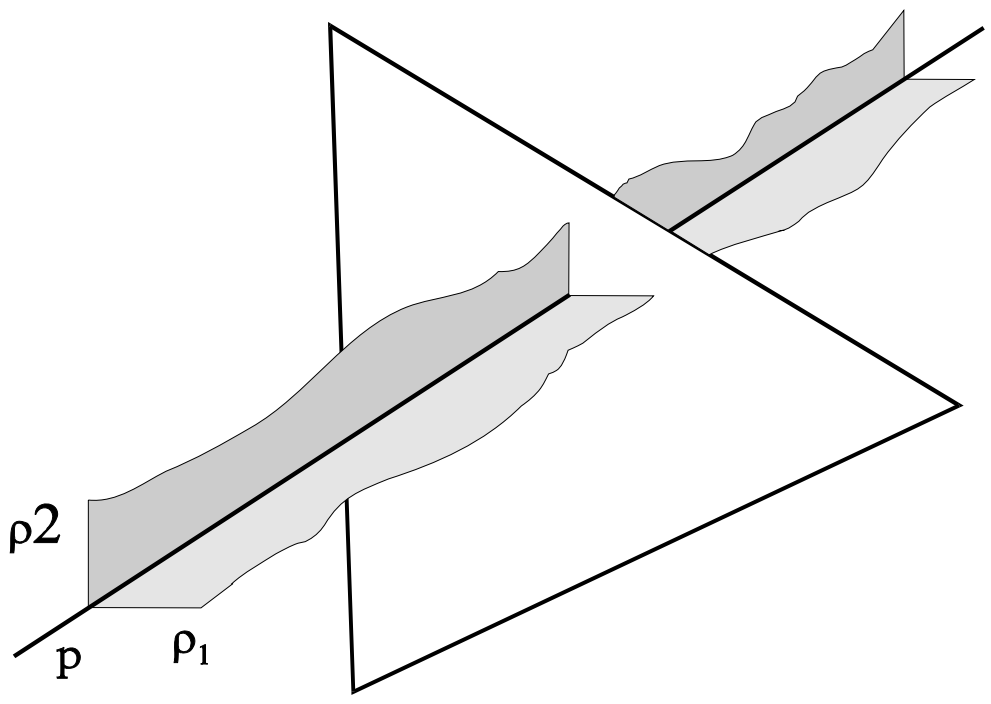

Usage of two planes for line definition
Figure 1.2

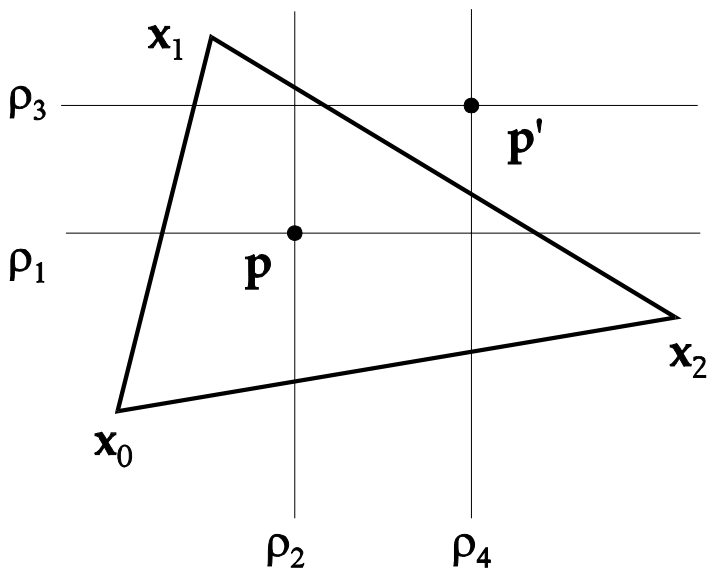

Possible kinds of intersection
Figure 1.3

It can be seen that any line *p* in E$^3$ can be defined as an intersection of two non-parallel planes $\rho_1$ and $\rho_2$, see Fig.1.2. It means that if the line *p* intersects the given facet then the planes $\rho_1$ and $\rho_2$ intersect the given facet, too, but **not vice versa** ($\rho_3$ and $\rho_4$), see Fig. 1.3.

It is possible to test all facets of the given polyhedron against $\rho_1$ and $\rho_2$ planes. If both planes intersect the given facet then **compute detailed intersection** test. The intersection of the given facet (triangle) with the $\rho_1$, resp. $\rho_2$ planes exists **if and only if** two vertices $\mathbf{x}_A$ and $\mathbf{x}_B$ of the facet exist such that

$$sign(F_i(x_A)) \neq sign(F_i(x_B))$$

where: $F_i(\mathbf{x}) = a_i\ x + b_i\ y + c_i\ z + d_i$ is a separation function for the i-th plane, i = 1, 2

$F_i(\mathbf{x}) = 0$ is an equation for the i-th plane $\rho_i$ .

Unfortunately there is some principal inefficiency in this approach as the separation function $F_1(\mathbf{x})$, resp. $F_2(\mathbf{x})$ are computed more times than needed because every vertex is shared by three facets (triangles) at least.

## 2. Principle of the $O(\sqrt{N})$ algorithm

Planes $\rho_1$ and $\rho_2$, that define the line *p* can be taken as parallel to arbitrary selected two axis, see Fig.1.2. If those planes are observed from the line *p* they are orthogonal. It can be seen that we get singular cases if the given line is parallel to one of the axes. Therefore we should find a criterion how to select planes $\rho_1$ and $\rho_2$ and how to avoid such singular cases, see Fig.2.1.

Let us suppose that we can generally select two planes $\rho_1$ and $\rho_2$, see Alg.2.1, from parallel planes $\rho_1^+$, $\rho_2^+$, $\rho_3^+$ , see Fig.2.1.

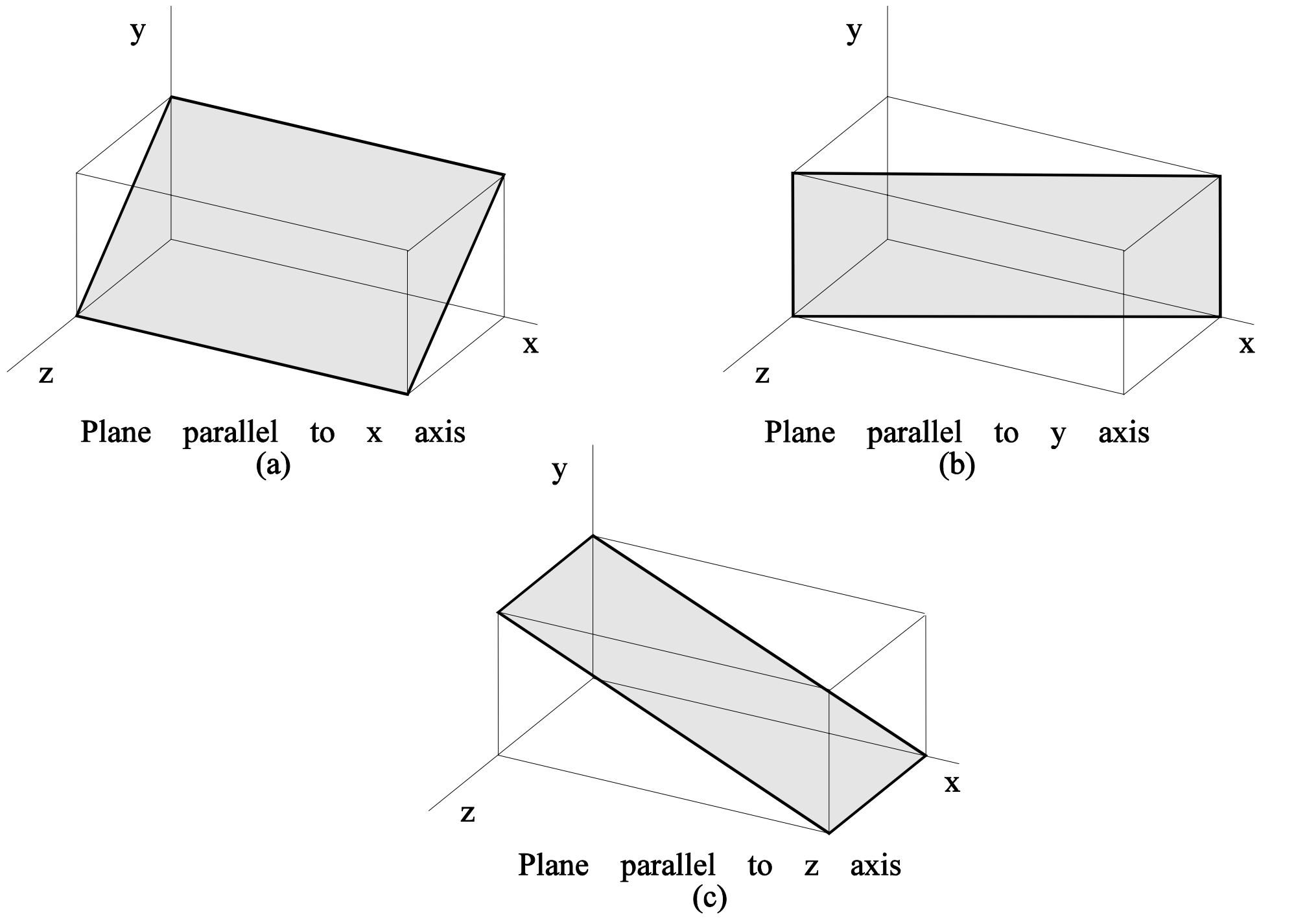

Definition of parallel planes
Figure 2.1

Those planes are defined as

$\rho_1^+ :\quad \mathbf{n}_1^T\mathbf{x} + d_1 = 0$ ,  where $\mathbf{n}_1 = \mathbf{s} \times \mathbf{e}_x$ , $e_x = [\,1, 0, 0\,]^T$

$\rho_2^+ :\quad \mathbf{n}_2^T\mathbf{x} + d_2 = 0$ ,  where $\mathbf{n}_2 = \mathbf{s} \times \mathbf{e}_y$ , $e_y = [\,0, 1, 0\,]^T$

$\rho_3^+ :\quad \mathbf{n}_3^T\mathbf{x} + d_3 = 0$ ,  where $\mathbf{n}_3 = \mathbf{s} \times \mathbf{e}_z$ , $e_z = [\,0, 0, 1\,]^T$

where $\mathbf{s} = [s_x,\ s_y,\ s_z]^T$ is a directional vector of the line *p*.

Planes $\rho_1$ and $\rho_2$ are selected so that

$$\rho_1 \; , \; \rho_2 \in \{ \rho_1^+ \; , \rho_2^+ \; , \rho_3^+ \}$$

see Alg.2.1.

ii := index of maximal value { | $s_x$ | , | $s_y$ | , | $s_z$ | };
**case** ii **of**
1: **begin** $\mathbf{n}_1 = \mathbf{s} \times \mathbf{e}_y$; $d_1 := -\mathbf{x}_1^T \mathbf{x}_A$;
$\mathbf{n}_2 = \mathbf{s} \times \mathbf{e}_z$; $d_2 := -\mathbf{x}_2^T \mathbf{x}_A$; { see Fig.2.1.a }
**end**;
2: **begin** $\mathbf{n}_1 = \mathbf{s} \times \mathbf{e}_z$; $d_1 := -\mathbf{x}_1^T \mathbf{x}_A$;
$\mathbf{n}_2 = \mathbf{s} \times \mathbf{e}_x$; $d_2 := -\mathbf{x}_2^T \mathbf{x}_A$; { see Fig.2.1.b }
**end**;
3: **begin** $\mathbf{n}_1 = \mathbf{s} \times \mathbf{e}_x$; $d_1 := -\mathbf{x}_1^T \mathbf{x}_A$;
$\mathbf{n}_2 = \mathbf{s} \times \mathbf{e}_y$; $d_2 := -\mathbf{x}_2^T \mathbf{x}_A$; { see Fig.2.1.c }
**end**;

Algorithm 2.1

We can project polyhedron to the xy plane, resp. yz or zx planes and we obtain two dimensional mesh of facets (triangular mesh). Generally we get two meshes - the front one and the back one. Now we can use an approach based on an idea to find sets of facets $\Omega_1$, resp. $\Omega_2$ intersected by planes $\rho_1$, resp. $\rho_2$. For intersection computation will be used only facets from the set $\Omega$ that is defined as:

$$\Omega = \Omega_1 \cap \Omega_2$$

It enables us to speed up the line clipping problem solution significantly. Using this approach we get faster algorithm with *O(N)* complexity or an algorithm with $O(\sqrt{N})$ complexity, see [Ska96a] and [Ska96c] for details.

## 3. Dual space representation

Any line $p \in E^2$ can be described by an equation

$$ax + by + c = 0$$

and rewritten as

$$y = kx + q \qquad \text{if} \qquad |k| \le 1 \qquad b \ne 0$$

resp.

$$x = my + p \qquad \text{if} \qquad |m| < 1 \; a \ne 0$$

It means, that the line $p \in E^2$ can be represented using an asymmetrical model of dual space representation as a point $D(p) = [k, q] \in D(E^2)$, resp. $D(p) = [m, p] \in D(E^2)$, see Fig.3.1. This representation model has very interesting properties and usage that can be found in [Sto89a], [Kol94a]. A similar approach can be used for other geometric primitives, like quadric surfaces etc. Generally it is possible to show relations between fundamental geometric primitives by the table Tab.3.1. In the following we will consider situations in $E^2$ only.

Komentář [VS1]:

| Space | Euclidean representation | Dual representation |
|---|---|---|
| $E^2$ | line<br>point | point<br>line |
| $E^3$ | plane<br>line<br>point | point<br>line<br>plane |

Fundamental relations between geometric primitives representations
Table 3.1.

It can be shown that a polygon $P \in E^2$, see Fig.3.1.a, can be represented by an infinite area in dual space $D(E^2)$ if asymmetrical model is used for dual space representation, see Fig.3.1.b, resp. Fig.3.1.c. There is also a possibility to use the symmetrical model, but it is not convenient for our considerations. For more theoretical background of dual space representation, see [Sto89a], [Zac96a].

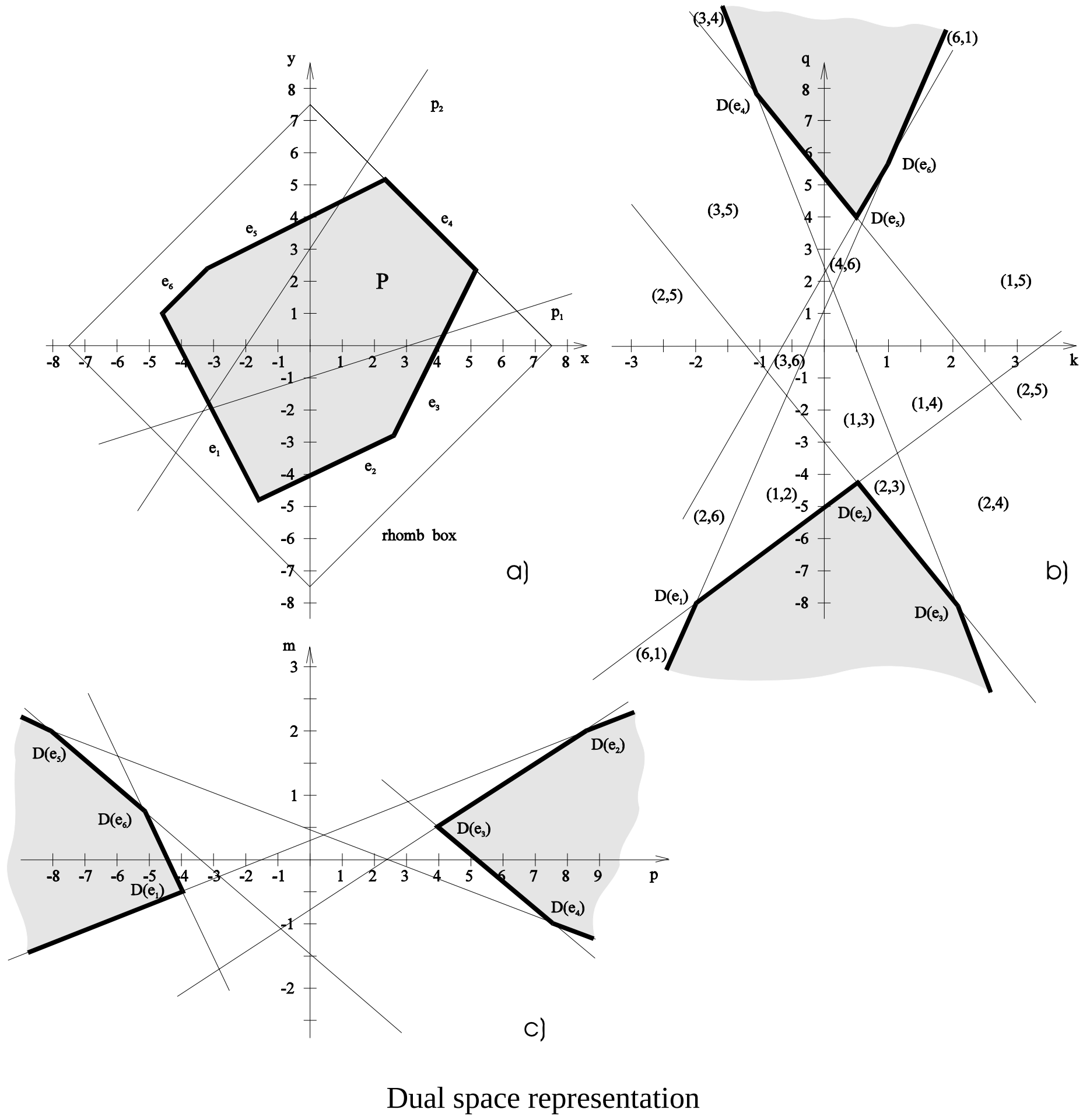

Dual space representation
Figure 3.1

The **test** whether a line $p \in E^2$ intersects a convex polygon $P \in E^2$ is dual to the well known Point-in-Polygon test. Algorithms for Point-in-Polygon test usually have *O(N)* or *O($lg_2$ N)* complexities. The algorithm complexity can be reduced to expected *O(1)* complexity if some pre-processing is used without application of parallel processing, see [Ska94b] for details.

Nevertheless the line clipping problem solution in $E^2$ generally consists of two steps:

- test whether a line intersects the polygon (dual to the Point-in-Polygon test),
- selection of polygon edges which are intersected by the given line and computation of intersection points.

It means that the line clipping problem solution is more complex than the Point-in-Polygon test.

It can be shown that the line *p* intersects edges *(a,b)* **if and only if** the point *D(p)* lies in the zone $(\alpha, \beta)$ in dual space representation, see Fig.3.1.

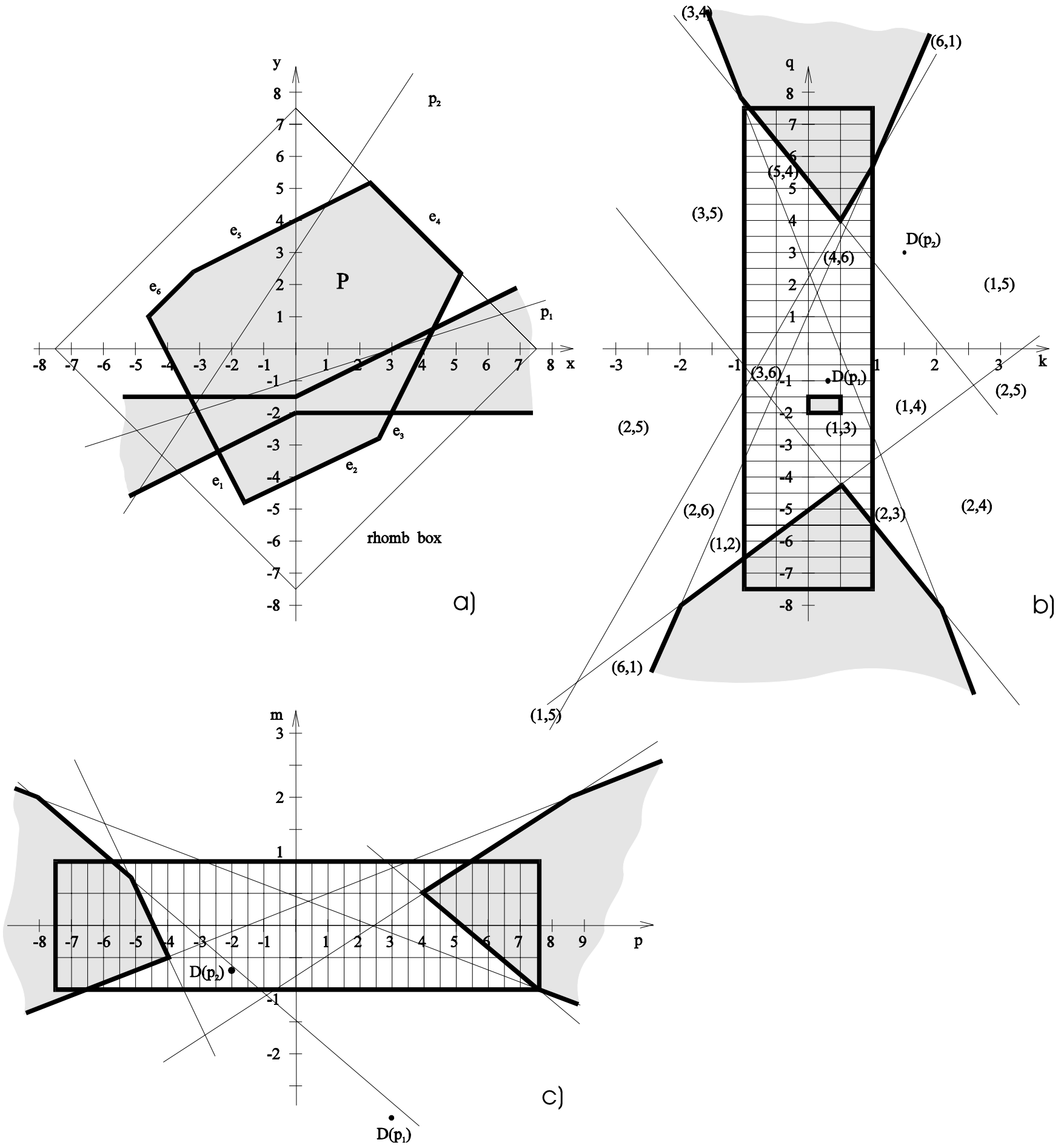

Semidual space representation
Figure 3.2

There are two problems if dual space representation is used that must be solved:

- dual space and zones are **infinite** and it is difficult to represent them,
- it is necessary to find fast method for zone determination in which zone the point *D(p)* lies.

Let us consider a modified rhomb box so that the given polygon is inside a rectangle, Fig.3.2.a. It can be seen that *q*, resp. *p* values are limited.
The given line $p \in E^2$ can be represented as

$$y = kx + q \qquad \text{if} \qquad |k| \le 1$$

and

$$x = my + pc \qquad \text{if} \qquad |m| < 1$$

If both representations are used then $k$, resp. $m$ values are limited. Then values $[k, q]$, resp. $[m, p]$ are from the limited area $< -1, 1 > x < -a, a >$ in both space representations. We will denote those two limited spaces as **semidual spaces**, see Fig.3.2.

Fast methods for zone detection in which a point *D(p)* lies are needed. One possibility is a usage of the space subdivision technique. If semidual spaces for $(k, q)$, resp. $(m, p)$ are subdivided into small rectangles then it is possible to pre-compute a list AEL (Active Edge List) of polygon edges that interfere in semidual space with the given rectangle. If rectangles are small enough then each AEL list will contain only two polygon edges. Each rectangle in the semidual space representation represents an infinite „butterfly"zone in $E^2$, see Fig.3.2.a.

It is necessary to point out that the rectangle that bounds the polygon must be as small as possible. Generally the limits for $q$ and *p* axis can differ. It decreases the memory requirements significantly.

## 4. Principle of the proposed algorithm

Let us assume that the given polyhedron $P \in E^3$ is projected to $E^2$, see Fig.4.1 (only the front facets are shown) and projected plane $\rho_1$ which is a line *p* after projection. This line intersects many facets (triangles), in our case the line $p$ intersects the facets 9,11,12,13,14,15,16,17,18,19. The plane $\rho_1$ described as

$$y = kx + q$$

(axis z is orthogonal to the xy plane) that is the line *p* in our projection.

Let us assume the semidual representation for *(k,q)* values. Then the semidual space can be split into small rectangles using space subdivision technique. Each rectangle in semidual space represents an infinite „butterfly" zone in $E^2$ space. There are AFL lists (Active Facets List) of facets associated with each „butterfly" zone. The AFL list contain information on all facets that interfere with the zone. The AFL list can be represented as a list of pointers but such an implementation would be quite memory demanding as its length can be estimated as $O(\sqrt{N})$. Because of that it is more convenient to use binary maps [Ska93b]. This technique is based on a binary vector in which the i-th bit is set to „1" if the i-th object is in the AFL list. Using this technique the memory requirements are small and the intersection operation is implemented as the bit-wise operation **and**.

Using this approach we can expect that 4 - 6 facets will be necessary to test in detail if semidual space is subdivided enough. It is necessary to point out that we must prepare both *(k,q)* and *(m,p)* semidual representations for all three planes $\rho_i^+$, $i = 1,2,3$, i.e. $AFL_1$,...,$AFL_j$ list. It means that we need six semidual representations altogether. For each clipped line $p$ we must select two planes $\rho_{i_1}$ and $\rho_{i_2}$ and appropriate semidual representations, i.e. *(k,q)* or *(m,p)* , for each selected plane. The proposed algorithm can be described by Alg.4.1.

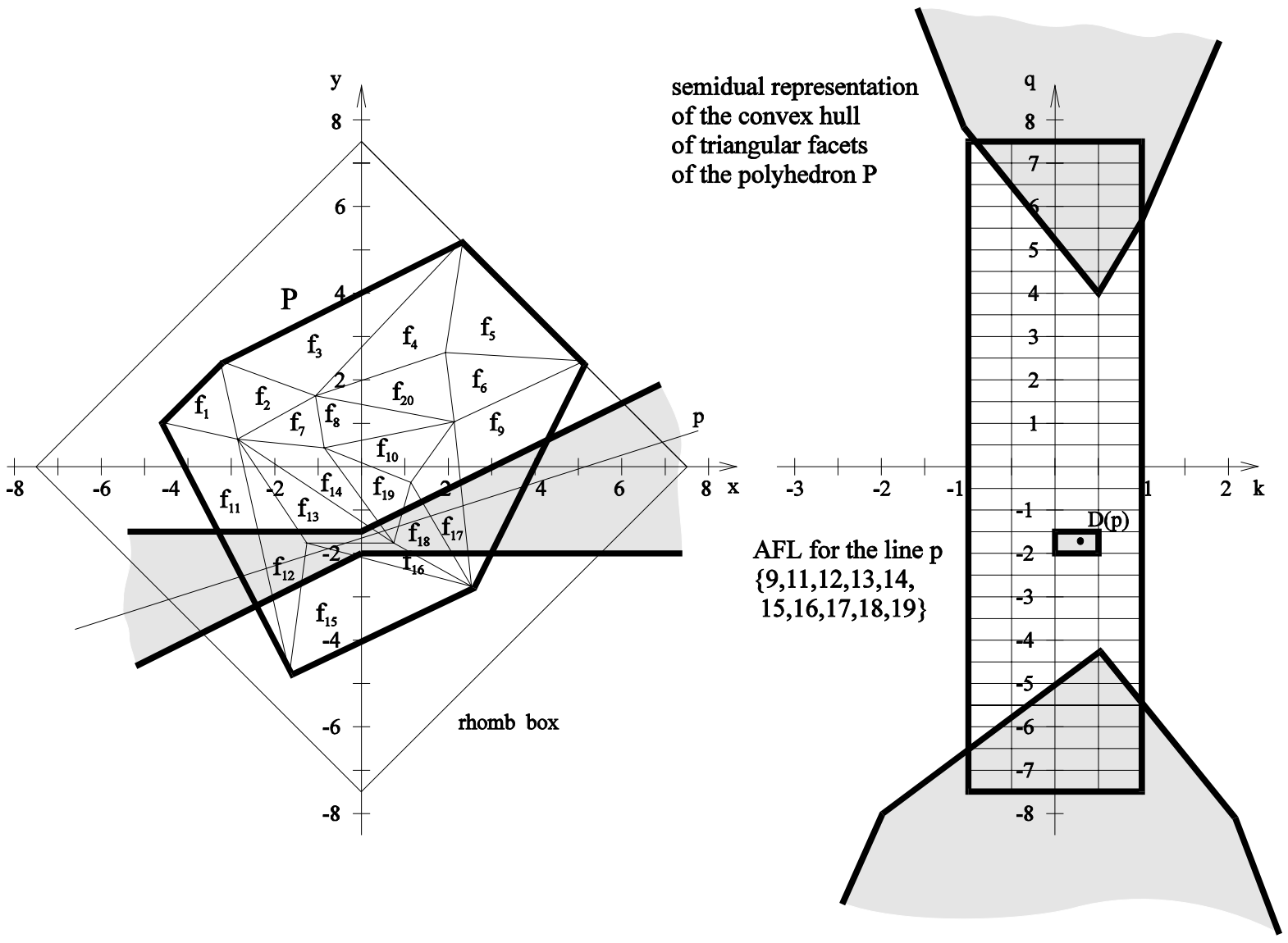

Figure 4.1

**global constants**: a - size of bounding rhomb box for the given polyhedron P,
$N_q$ - number of subdivision for q axis in semidual space representation,
$N_k$ - number of subdivision for k axis in semidual space representation,
for all other spaces assume $N_q = N_p$ , $N_k = N_m$ .

Select two planes $\rho_{i_1}$ and $\rho_{i_2}$ , $i_1 \neq i_2$ for the given line $p \in E^3$, according to Alg.2.1;
k := 1;
**for** i := $i_1$, $i_2$ **do** (* plane index i $\in$ {1,2,3} *)
**begin** COMPUTE line equation of $p_i$; (* projected plane $\rho_i$ *)
(* $ax + by + c = 0$ *)
**if** $|\Delta x| \geq |\Delta y|$ **then** j := 2*i - 1 **else** j := 2*i; (* index of the $AFL_j$ list *)
(* $AFL_1$ is the (k,q) semidual space for xy plane *)
(* $AFL_2$ is the (m,p) semidual space for xy plane, etc. *)
ii := 2*a / $N_q$ * q; jj := 2 / $N_k$ * k; (* index zone determination *)
$\Omega_k$ := $AFL_j$[ii,jj]; k := k + 1;
**end**;
**$\Omega := \Omega_1 \cap \Omega_2$**;
**for** i := 1 **to** N **do** (* N - number of polyhedron facets *)
**if** $\Omega_i$ = 1 **then** Detail $E^3$ Test (facet$_i$, p); (* computation is done for 4 - 6 facets only *)

Algorithm 4.1

Function *Detail $E^3$ Test* is based on the CB algorithm that is performed only for those facets that are included in the final set Ω. It is obvious that the algorithm complexity does not depend on the number of polyhedron facets but on the length of the set Ω. If the rectangles are small enough then 4 - 6 facets can be expected in the final set Ω nearly for all cases. Because all steps in Alg.4.1 have *O(1)* complexity the whole algorithm has *O(1)* complexity, too. It is necessary to point out that number of members in AFL list depends on subdivision in *(k,q)*, resp. *(m,p)* spaces and also on geometric shape of the given polyhedron, see [Ska94b].

## 5. Construction of AFL list

An algorithm for setting the AFL list directly is quite complicated. A simple solution how to set up the $AFL_1$ lists for all zones in *(k,q)* semidual space is described by Alg.5.1.

```
for i:=1 to N_k-1 do
    for j:=1 to N_q-1 do
        for k:=1 to N do
            if facet_k interferes with the zone (i,j) defined
                              by corners (i,j) and (i+1,j+1)
                then add facet_k into the AFL_1[i,j] list;
```

Algorithm 6.1

Because number of facets $N$ is reasonably small and there are no special cases this method is fast enough. The AFL list for *(m,p)* semidual space can be determined in a similar way. This step is necessary repeat for all other four cases and we get $AFL_3$ - $AFL_6$ lists.

Now it is necessary to find a criterion how to choose the $N_k$ and $N_q$ , resp. $N_m$ and $N_p$ values. It can be shown that for *(k,q)* semidual space it is necessary to calculate $[k,q]$ values for all polyhedron edges from the equation

$$y = kx + q$$

if xy plane is considered. Then

$N_q > 2a / \Delta y$ where $\Delta y = \min \{| y_i - y_j |\}$ for all $i, j$ & $i \neq j$ & $y_i \neq y_j$

$N_k > 2 / \Delta k$ where $\Delta k = \min \{| k_i - k_j |\}|$ for all $i, j$ & $i \neq j$ & $k_i \neq k_j$

Similarly for $(m, p)$ semidual space

$$x = my + p$$

and

$N_p > 2a / \Delta x$ where $\Delta x = \min \{| x_i - x_j |\}$ for all $i, j$ & $i \neq j$ & $x_i \neq x_j$

$N_m > 2 / \Delta m$ where $\Delta m = \min \{| m_i - m_j |\}$ for all $i, j$ & $i \neq j$ & $m_i \neq m_j$

It means that the $N_k$ and $N_q$ , resp. $N_m$ and $N_p$ values depend on geometric shape of the given polyhedron *P*. For detailed description, see [Ska96b].

## 6. Theoretical considerations and experimental results

The proposed algorithm has been tested and compared with the CB algorithm as the CB algorithm is very stable and its behaviour more or less does not depend on geometric properties of the given polyhedron and on clipped lines. Because the proposed algorithm is supposed to be superior over other modifications of CB algorithm it is necessary to make theoretical estimation of its efficiency. It is necessary to point out that algorithm efficiency can differ from computer to computer. For used PC 386DX/33 MHz we got for $5.10^7$ operations ( := , < , ± , * , / ) the following timing ( 33 , 50 , 16 , 20 ,114 ).

Let us assume that $N$ is a number of facets of the given polyhedron. For algorithm efficiency considerations we will consider:

- CB algorithm complexity, see [Ska93a], can be described as

$$T_{CB} = (6,3,6,6,1) * N$$

that is for considered timing

$$T_{CB} = 777 * N$$

- proposed *O(1)* with complexity defined as

$$T_{O(1)} = (18,3,8,8,4) + T_{CB}(2)$$

and using considered timing

$$T_{O(1)} = 3042$$

Let us introduce algorithm efficiency coefficients as:

$$\nu_1 = \frac{T_{CB}}{T_{O(1)}}$$

then the expected efficiency of the proposed algorithm is

$$\nu_1 = \frac{T_{CB}}{T_{O(1)}} = \frac{777 * N}{3042} \approx 0.25 * N$$

It means that for $N \geq 1000$ we get significant speed up if processing time is considerd only. For experimental results see fig.6.1. It shows total time including preprocessing for a polyhedra with different number of facets for 1000 clipped lines.

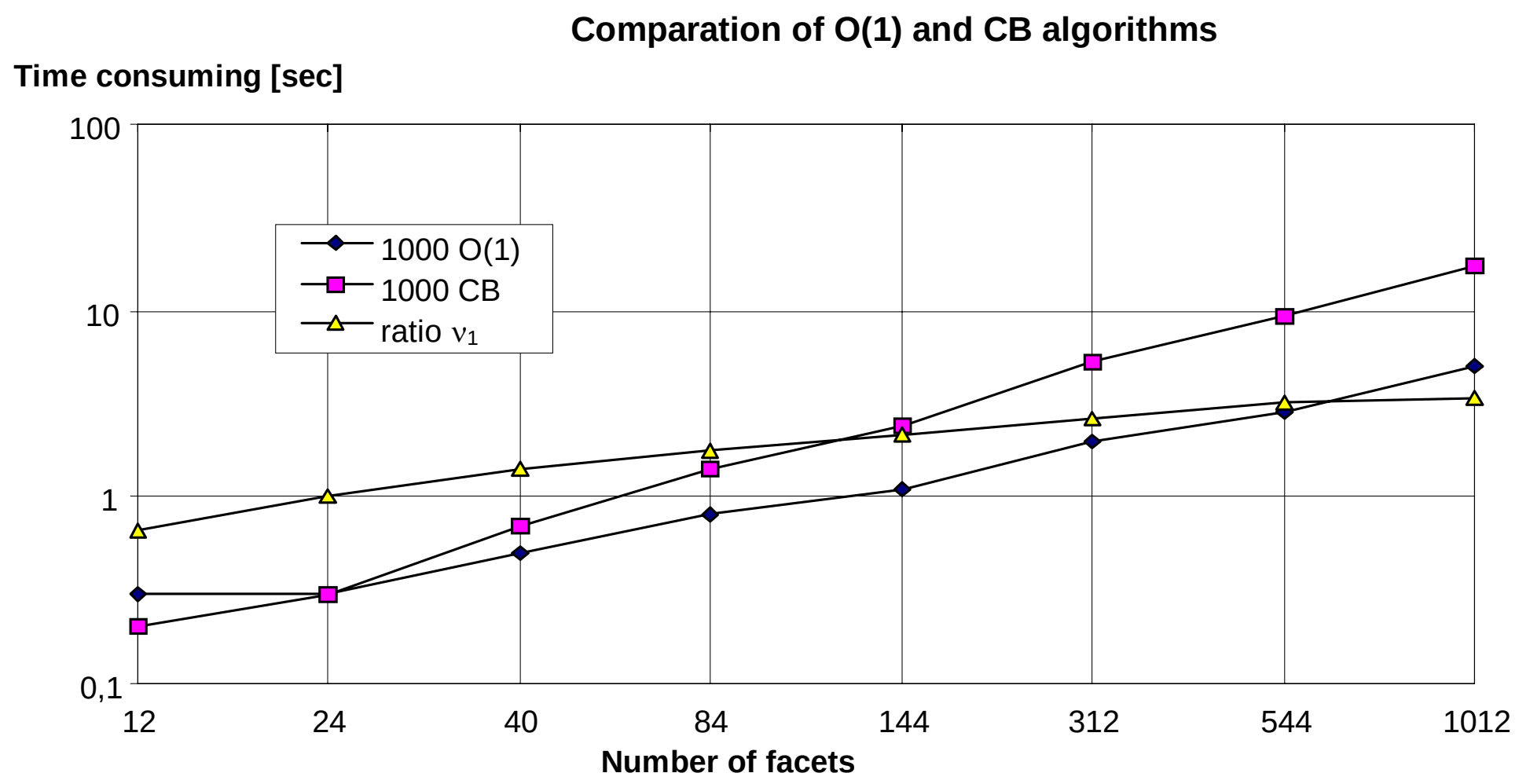

Figure 6.1.

How total processing time (including preprocessing) depends on subdivision of *k* and *q* is shown at fig.6.2 for a polyhedra represented by 2112 facets.

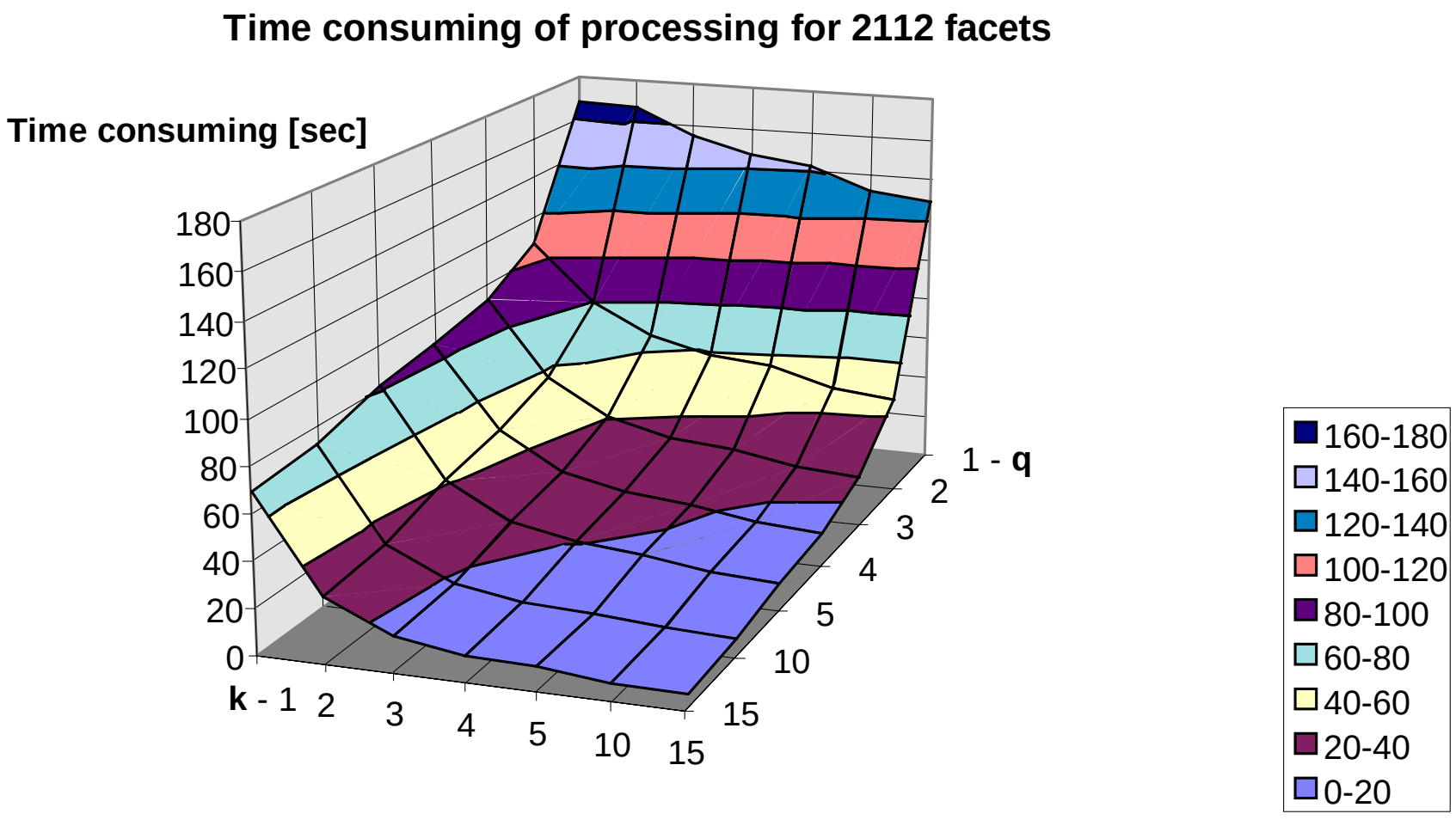

Figure 6.2

Fig.6.3. shows how number of facets in one AFL list (before intersection operation) depends on subdivision in *k* and *q*.

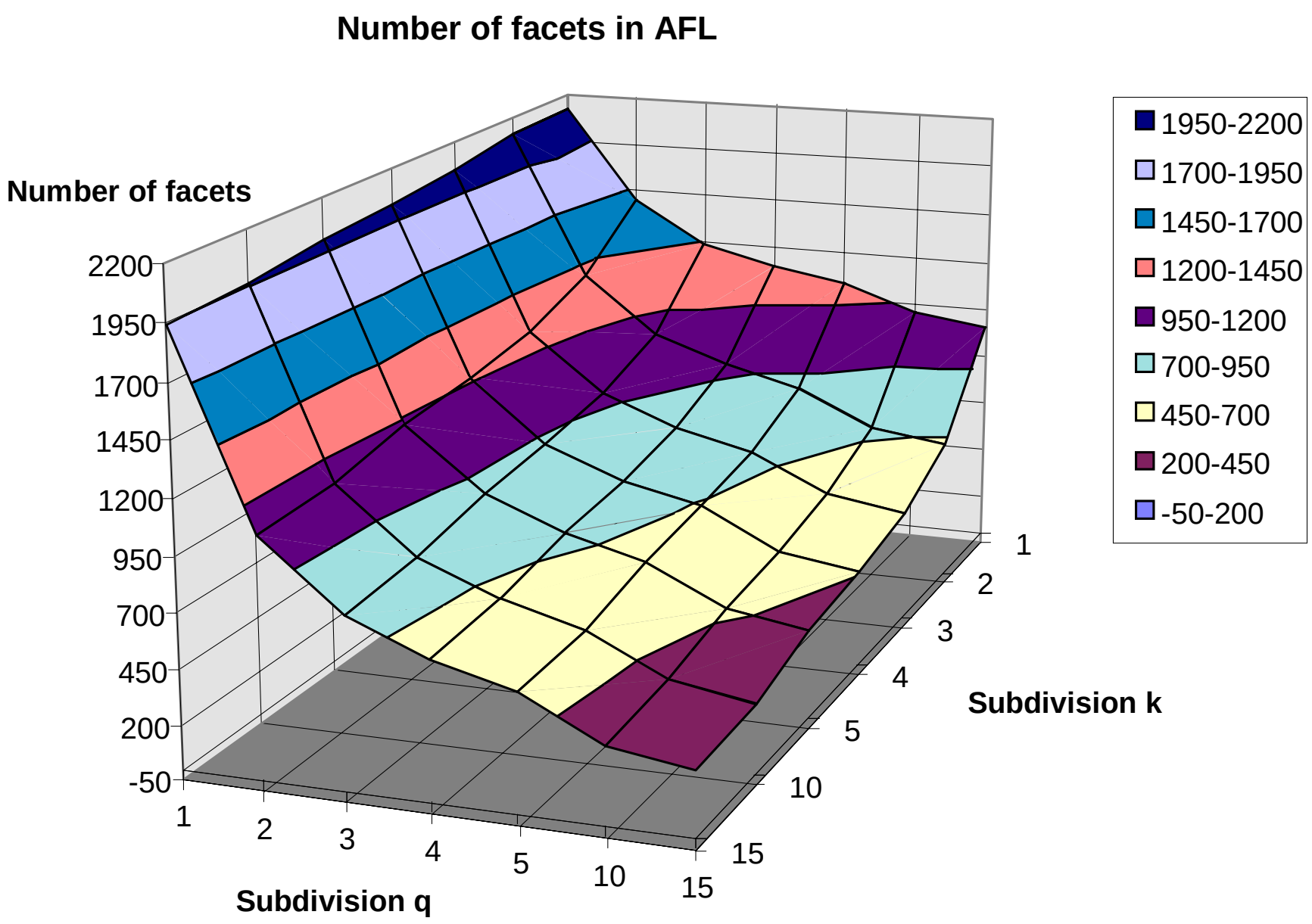

Figure 6.3.

For detailed comparative study see [Ska96d].

## 7. Conclusion

The new line clipping algorithm against convex polyhedron in $E^3$ was developed. The proposed algorithm is convenient for those applications where clipping area is constant and many lines are clipped. The algorithm claims the processing complexity *O(1)* and superiority over the CB algorithm modifications with *O(N)* complexity.

The presented approach can be applied in many areas of computer graphics and there is a hope that it can be used in ray tracing techniques, too. Some papers can be downloaded using the author's http address.

## 8. Acknowledgements

The author would like to express his thanks to Mr.B.Sup for careful implementation and testing algorithms and to all who contributed to this work, especially to recent MSc. and PhD. students of Computer Graphics courses at the Univ. of West Bohemia in Plzen who stimulated this work and for many suggestions they proposed.